\documentclass[pra,aps,tightenlines,twocolumn]{revtex4}
\usepackage[]{graphicx}
\usepackage[]{epsfig}
\begin{document}

\title{Universal Optical Frequency Comb}

\author{A. A. Savchenkov, A. B. Matsko, W. Liang, V. S. Ilchenko, D. Seidel, and L. Maleki}

\affiliation{OEwaves Inc., 2555 E. Colorado Blvd., Ste. 400, Pasadena, California 91107}

\date{\today}

\begin{abstract}
We demonstrate that whispering gallery mode resonators can be utilized to generate optical frequency combs based on four wave mixing process at virtually any frequency that lies in the transparency window of the resonator host material.  We show theoretically how the morphology of the resonator can be engineered to produce a family of spectrally equidistant modes with anomalous group velocity dispersion appropriate for the comb generation.  We present experimental results for a frequency comb centered at 794~nm to support our theoretical findings.
\end{abstract}

\maketitle

Since their introduction over a decade ago, optical frequency combs have revolutionized many applications across the fields of science and metrology, and have opened new vistas in sensing and spectroscopy.  The conventional approach for optical comb generation with femtosecond lasers was recently augmented with optical combs generated using whispering gallery mode (WGM) resonators made of materials possessing Kerr nonlinearity \cite{delhaye07n,delhaye08prl,savchenkov08prl,grudinin09ol,agha09oe,levy09np,razzari09np,braje09prl,delhaye09arch,arcizet09chapt,matsko09sfsm,chembo10prl}.  These small structures address one of the shortfalls of femtosecond laser frequency combs, namely their size and relatively complex architectures that limits their range of applicability outside research laboratories, and thus restricts their applications.   The second shortfall of optical frequency combs related to the frequency of operation and has not been effectively addressed.  Though a number of optical combs covering selected spectral range from UV to IR have been demonstrated, there is no universal approach that can provide an optical frequency comb centered at any desired wavelength. We propose a solution of the problem in this paper.

The generation of a phase locked frequency comb with an optically pumped WGM microresonator is mediated by hyper-parametric oscillation (modulational instability \cite{haelterman92ol,coen97prl,coen01ol}) resulting from four wave mixing (FWM), and by group velocity dispersion. In the FWM process, two pump photons and two sideband photons are involved.  The generation of two sidebands symmetrically around the excited mode starts when the power of a continuous wave laser pumping the mode exceeds some threshold value \cite{matsko05pra}. Increasing the pump power results in generation of multiple sidebands via the hyper-parametric process, producing a phase locked group of tines in the optical comb. Depending on the dispersion of the WGM spectrum, the initial generation of sidebands separated by multiple free spectral range (FSR) could have a lower threshold compared with that of closely separated sidebands \cite{savchenkov08prl,chembo10prl}.

Group velocity dispersion (GVD) also plays a critical role in the generation of frequency combs. To explain this, let us recall that the threshold condition for the average number of photons, $\bar N$, in the pumping mode necessary to start the FWM oscillation process, which generates the Kerr frequency comb in a WGM resonator, reads as \cite{matsko05pra}
\begin{equation} \label{threshold}
4(g^2\bar N^2-\gamma^2) = \left [ 2(\omega_0-\omega)+(\omega_++\omega_--2\omega_0)-4 g \bar N \right ]^2,
\end{equation}
where $\gamma$ is the full width at the half maximum of the optical modes (we assume that the modes are identical in loss), $\omega_0$ and $\omega_{\pm}$ are the frequencies of the optically pumped mode and the first pair of the modes where FWM sidebands are generated at, $\omega$ is the carrier frequency of the external monochromatic pump, $g$ is the nonlinearity parameter
\begin{equation}\label{g}
g= \omega_0 \frac{\hbar \omega_0 c}{{\cal V} n_0} \frac{n_2}{n_0},
\end{equation}
$c$ is the speed of light in the vacuum, ${\cal V}$ is the mode volume, $n_0$ and $n_2$ are the linear and nonlinear refractive indices of the resonator's host material.

In accordance with Eq.~(\ref{threshold})  oscillation is allowed if at least one of the following two conditions is satisfied: i) the optical pump is detuned to the lower frequency than the corresponding optical resonance (red detuned), $\omega_0 >\omega$; and ii) the resonator modes are characterized with anomalous dispersion, $\beta_2 <0$  ($2\omega_0-\omega_+-\omega_-\simeq c\beta_2 \omega_{FSR}^2/n_0$, where $\omega_{FSR}$ is the free spectral range of the resonator). The condition (ii) is usually stronger than (i). Even though oscillation can occur in a resonator possessing arbitrary frequency dispersion in the mode family participating in the comb generation \cite{haelterman92ol,coen97prl,coen01ol}, it generally starts when the dispersion is small and anomalous. Resonators with normal dispersion exhibit much weaker FWM process, which frequently competes with stimulated Raman scattering (SRS), which has a lower threshold for oscillation \cite{spillane02n}. The red detuning of the pump light with respect to the corresponding resonator mode is required for phase matching and compensation of the dispersion, though it  increases the oscillation threshold power, so the SRS process starts first. By contrast, anomalous dispersion simplifies the phase matching of the FWM process \cite{agha09oe} and removes the burden away form laser detuning. These  assertions have been confirmed experimentally. For instance, for combs observed with WGM resonators, a value of $(2\omega_0-\omega_+-\omega_-)/(2\pi) \simeq -2.6$~MHz  was measured for a 75~$\mu$m fused silica microresonator  at 1,550~nm \cite{delhaye07n}. The dispersion of a 0.7~cm CaF$_2$ resonator was $(2\omega_0-\omega_+-\omega_-)/(2\pi) \simeq -300$~Hz \cite{savchenkov08oe}, and a 0.255~cm CaF$_2$ resonator had $(2\omega_0-\omega_+-\omega_-)/(2\pi) \simeq -680$~Hz \cite{savchenkov08prl}.

The dispersion $\beta_2$ includes both geometrical and material parts. The geometrical dispersion is usually normal. To obtain mode spectra characterized with small anomalous dispersion one needs to produce a resonator out of a material with anomalous dispersion for the given laser wavelength. The shape and size of the resonator must be properly selected to have an overall small and anomalous dispersion ($|2\omega_0-\omega_+-\omega_-| \ll \gamma $). These requirements prohibit generation of  Kerr frequency combs with an arbitrary repetition rate at an arbitrary wavelength, since the repetition rate is determined by the resonator size, and the wavelength is determined by the material dispersion.  For instance, a fused silica resonator is not expected to generate Kerr frequency comb if pumped at a wavelenght shorter than 1.3~$\mu$m.

Now, consider a spheroidal resonator with two nearly equidistant mode sequences characterized with frequency intervals (free spectral range, FSR) $FSR_l=c/(2\pi n a)$ and $FSR_p=FSR_l(a-b)/b$, where $l$ and $p$ are the azimuthal and the vertical numbers of the mode, $c$ is the speed of light in the vacuum, $n$ is the refractive index of the material, and $a$ and $b$ are the semi-axes of the spheroid \cite{gorodetskyo6jstqe}. Modes belonging to a basic sequence characterized with  $FSR_l$ are generally utilized for generation of Kerr combs. However, for the purpose of comb generation at any arbitrary frequency within the transparency window of the resonator material, we propose to use the vertical mode family characterized with  $FSR_p$, which depending on the ratio $a/b$ can have an FSR significantly different from $FSR_l$. An important feature of a vertical mode family is that the mode spacing strongly depends on the shape of the spheroid, and the corresponding GVD can be anamoulous.

\begin{figure}[ht]
\centerline{\epsfig{file=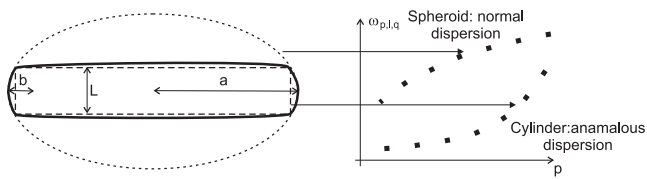,width=8.5cm,angle=0}}
\caption{\label{fig1}  Illustration of the cross sections of a spheroid and a cylinder, and a WGM resonator being a convolution of those shapes. A spheroidal WGM resonator has normal geometrical group velocity dispersion in vertical modes, and nearly equidistant spectra. A cylindrical resonator has anomalous group velocity dispersion for its modes but not equidistant spectra. The resonator designed as a convolution of those two shapes has  a nearly equidistant spectrum and anomalous group velocity dispersion necessary for achieving phase matching in the FWM process for any resonator host material. }
\end{figure}
The morphological dependence of the dispersion can be understood as follows. Consider the convolution of a cylindrical and a spheroidal resonator (Fig.~\ref{fig1}). The frequency spectrum of the cylinder is given by
\begin{equation}
\omega_{p,l,q} = \frac {c} {n} \sqrt{k_{l,q}^2+\left ( \frac{p \pi}{L} \right )^2},
\end{equation}
where $k_{l,q} a \approx l + \alpha_q (l/2)^{1/3}$ is the azimuthal wave number given by the spherical Bessel functions, $q$ is the radial wave number of the mode, $\alpha_q$ is q$^{th}$ root of Airy function ($Ai(-\alpha_q)=0$), and $L$ is the hight of the cylinder. In the case of $l \gg p$ the spectrum is significantly non-equidistant with $p$, but the group velocity dispersion is anomalous $2\omega_{p,l,q}-\omega_{p+1,l,q}-\omega_{p-1,l,q} <0$. The nearly equidistant frequency spectrum of a spheroid can be approximated by \cite{gorodetskyo6jstqe}
\begin{equation}
\omega_{p,l,q} \simeq \frac {c} {n}  \left [ k_{l,q} + \frac{(a-b)(1+2p) }{2 a b} - \frac{a p^2}{2b^2l} \right ],
\end{equation}
so that the dispersion is normal, $2\omega_{p,l,q}-\omega_{p+1,l,q}-\omega_{p-1,l,q} > 0$. A resonator being a convolution of the cylinder and spheroid has a frequency spectrum given by
\begin{equation}
\omega_{p,l,q} \simeq \frac {c} {n} \left [ k_{l,q} + \frac{(a-b)(1+2p) }{2ab} + \xi p^2 \right ],
\end{equation}
where $\xi$ is a numeric parameter depending on $L$, $a$, and $b$. The parameter $\xi$ can be positive in the case of comparably small $L<2b$, which means that the dispersion of the vertical resonator modes is anomalous. In this way, the anomalous dispersion can be used to compensate for the normal material dispersion in the Kerr comb.

There are no measurements of the group velocity dispersion for the majority of optical materials, and rigorous calculations confirming these prediction are rather mathematically involved. So we performed an experiment to verify these estimates. We designed and fabricated a truncated spheroidal crystalline CaF$_2$ WGM resonator and pumped it with 794~nm light emitted by a semiconductor laser that was self-injection locked \cite{wei10ol} to a selected WGM. The light was sent in and retrieved out of the resonator using a glass coupling prism. The exiting light was collimated and injected into a single mode optical fiber to be further analyzed. The spheroidal resonator with $3$~mm and $3/4$~mm semi-axes (with $a/b \approx 4$) was fabricated out of a z-cut CaF$_2$ wafer by mechanical polishing. The loaded quality (Q-) factor of the resonator exceeded $3\times10^9$ ($126$~kHz full width at the half maximum of the corresponding WGMs).

\begin{figure}[ht]
\centerline{\epsfig{file=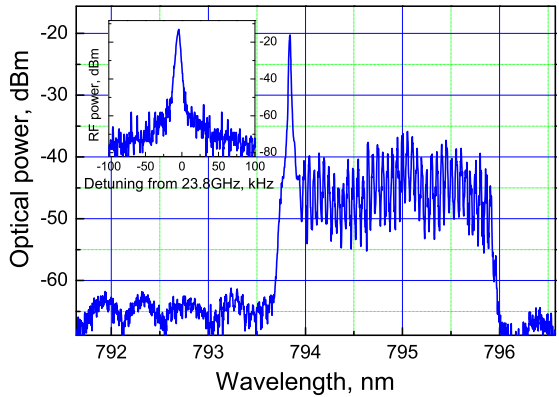,width=8.5cm,angle=0}}
\caption{\label{fig2}  The output spectra of a critically coupled CaF$_2$ resonator pumped with 2~mW of 794~nm light. No stimulated Raman scattering is observed. The resonator has 8~GHz free spectral range, while the optical frequency comb has 23.78~GHz repetition rate. The comb harmonics are phase locked, so they generate a coherent RF signal (inset) on a fast photodiode.}
\end{figure}

We observed the onset of hyperparametric oscillation with 0.1~mW of optical power. Increasing  the power to 2~mW resulted in the formation of a well pronounced optical frequency comb shown in Fig.~(\ref{fig2}). The frequency envelope of the comb suggests that the vertical mode family is involved in this process. The asymmetry of the spectrum results from the significant spatial spread of the light emitted from the resonator that is selectively collected with the fiber. Since the group velocity dispersion of the material is normal at this wavelength ($\beta_2 =28$~ps$^2/$km) and the difference between two consecutive FSR's is positive and large $(2\omega_0-\omega_+-\omega_-)/(2\pi) \simeq 21$~kHz, but the comb is still generated, we conclude that the anomalous dispersion of the resonator compensated the material dispersion. The experiment proves our theoretical prediction.

Note that the repetition frequency of the observed comb is approximately three times as large as the basic mode frequency FSR. In our case $a/b \approx 4$, so that $FSR_p\approx 3 FSR_l$, which is close to experimentally measured value $FSR_p/FSR_l=23.78\ GHz/8\ GHz=2.96$, where 23.78~GHz is the repetition frequency of the comb we observed. This is further proof that the frequency components of the Kerr comb reported here are not generated in the same family of fundamental modes with different  orbital momenta. Rather, the components belong to the family of modes with different vertical quantum number $p$ and the same orbital momentum $l$.

The maximum number of Stokes and anti-Stokes components of a conventional Kerr comb generated in the basic mode family \cite{delhaye07n,delhaye08prl,savchenkov08prl,grudinin09ol,agha09oe,levy09np,razzari09np,braje09prl,delhaye09arch,arcizet09chapt,matsko09sfsm,chembo10prl} is limited by the dispersion and nonlinearity of the resonator and can be very large.  On the contrary, the number of the Stokes comb components observed in our system is limited by the vertical index $p$ of the mode that is selected for optical excitation. Indeed, figure (\ref{fig2}) represents an optical comb produced by pumping the mode with $p=42$. The saturated comb has a flat top and a sharp edge at the lower frequency side, because there are simply no modes with negative transverse index. This type of Kerr comb affords much more flexibility compared with a comb generated at the basic mode sequence: its repetition frequency is large in an oblate spheroid, as observed in our experiment, and a nearly spherical resonator would generate a comb with repetition frequency much lower than the azimuthal FSR of the resonator. In general, the repetition frequency of a transverse comb depends symmetrically on both the resonator semi-axes $a$ and $b$
\begin{equation}
\nu_{comb}=\frac{c}{\pi n} \left |\frac{1}{b}-\frac{1}{a}  \right |.
\end{equation}
This property results in the ability for generation of low frequency combs even in very small microresonators of very low mode volume, and therefore, very low threshold. It also ultimately leads to a new kind of RF photonic oscillator \cite{savchenkov04prl} with extremely low phase noise.

The asymmetry of the observed comb envelop requires a special explanation. A WGM of vertical index $p$ is described by an eigenfunction which has $p+1$ maxima in the electric field along the vertical axis of the resonator \cite{gorodetskyo6jstqe}. The overlap of the evanescent field of the mode on the coupling prism determines the emission pattern. The emission pattern of the transverse mode consists of two symmetrical lobes with respect to the horizontal plane.  Since the angular space between the lobes and the frequency of the comb components grows with index $p$, components of lower index $p$ are concentrated closer to the equatorial plane of the resonator. The optical fiber coupler that we use is aligned in the equatorial plane of the resonator and thus has much higher efficiency for collecting the modes with smaller $p$ and, therefore, lower frequency.  This is why the comb as observed has the appearance of being asymmetrical.  We note that the spatial pattern of the comb emission provides a simple way to physically separate strongly correlated Stokes and anti-Stokes components of the comb for applications where correlated photons are desired.

In conclusion, we have proposed and experimentally validated a universal method for the generation of Kerr frequency combs at any desirable wavelength, and with any desirable repetition rate. The method is based on the management of the geometrical group velocity dispersion of whispering gallery modes participating in the process through the proper engineering of the shape of the resonator. We have found that the vertical sequence of the modes belonging to a specially shaped resonator has nearly equidistant spectrum characterized with anomalous group velocity dispersion. Using these modes we were able to generate a frequency comb at 794~nm center wavelength in a properly designed calcium fluoride resonator. The conventional Kerr comb produced with the basic sequence of the azimuthal modes was absent in our observations.



\end{document}